\documentclass[manuscript]{acmart}

\usepackage{adjustbox}
\RequirePackage{fix-cm}
\usepackage{graphicx}
\usepackage{wrapfig}
\usepackage{subcaption}
\usepackage{xcolor}
\newcommand{\CHANGED}[1]{\textcolor{blue}{#1}}

\newcommand{\respto}[1]{
\fcolorbox{black}{black!15}{%
\label{resp:#1}%
\bf\scriptsize R{#1}} }

\usepackage{color,colortbl}
\usepackage{url}
\usepackage{xcolor}
\usepackage{hyperref}
\usepackage{ifthen}

\newboolean{showcomments}
\setboolean{showcomments}{true} 
\ifthenelse{\boolean{showcomments}}
    {
        \newcommand{\response}[2]{\textcolor{blue}{{\it \\ \noindent \textbf{Response #1}: #2 \\ }}}
        
        \newcommand{\neil}[1]{\textcolor{olive}{{\it 
        	[Neil says: #1]}}}
        \newcommand{\ben}[1]{\textcolor{violet}{{\it [Ben says: #1]}}}
        \newcommand{\teresa}[1]{\textcolor{green}{{\it [MTB says: #1]}}}        
    }
    {
        \newcommand{\response}[1]{}
        \newcommand{\tim}[1]{}
        \newcommand{\neil}[1]{}
        \newcommand{\ben}[1]{}
          \newcommand{\teresa}[1]{}
    }
\usepackage{hyperref} 
\begin{document}


\title{Crowdsourcing the State of the Art(ifacts)}



\author{Maria Teresa Baldassarre}
\orcid{0000-0001-8589-2850}
 \affiliation{University of Bari, Italy}
 \email{mariateresa.baldassarre@uniba.it} 
\author{Neil Ernst}
\orcid{0000-0001-5992-2366}
 \affiliation{University of Victoria, Canada}
 \email{nernst@uvic.ca}
\author{Ben Hermann}
\orcid{0000-0001-9848-2017}
 \affiliation{Technische Universität Dortmund, Germany}
 \email{ben.hermann@cs.tu-dortmund.de}
\author{Tim Menzies, Rahul Yedida }
\orcid{0000-0002-5040-3196}
 \affiliation{NC State University, USA}
 \email{timm@ieee.org. ryedida@ncsu.edu}

\date{Received: date / Accepted: date}

\newcommand{\bi}{\begin{itemize}}
\newcommand{\ei}{\end{itemize}}

\begin{abstract} 
In any field,
finding   the ``leading edge'' of 
research   is
an on-going challenge. 
Researchers cannot  
appease reviewers  and
educators cannot teach to the leading edge of their field
if no one agrees on what is the state-of-the-art.

Using a novel crowdsourced
``reuse graph'' approach, we   propose here a
new method to  learn this state-of-the-art.
Our  reuse graphs are  less effort to build and  verify than other  community
monitoring methods (e.g. artifact tracks or citation-based  searches).
Based on a study of  170 papers from software engineering (SE) conferences in 2020, we have found over 1,600  instances of  reuse; i.e.,
reuse is rampant in SE research. Prior pessimism about a lack of reuse
in SE research
may have been a result of using the wrong methods to measure
the wrong things.


\end{abstract}
 \maketitle

\section{Introduction}
\label{intro}

According to Popper~\cite{popper2014conjectures}, the ideas we can {\em most trust} are
those that have been {\em most tried} and {\em most tested}. For that reason,
many of us are involved in the process called 
``Science''  that produces trusted knowledge by sharing one's ideas, and trying out and testing others' ideas.
Science and scientists form
communities where people do  each other
the courtesy of curating, clarifying, critiquing and improving
a large pool of  ideas. 

Prior to this study, the standard conclusion
was that researchers in the field of software engineering are rarely reusing research results (e.g., da Silva et al. reported that from 1994-2010, only 72 studies had been replicated by 96 new studies \cite{daSilva2012}).
If true, this is a significant problem since
not knowing the state of the art complicates
both research
and graduate education.

\begin{figure}
\includegraphics[width=\textwidth]{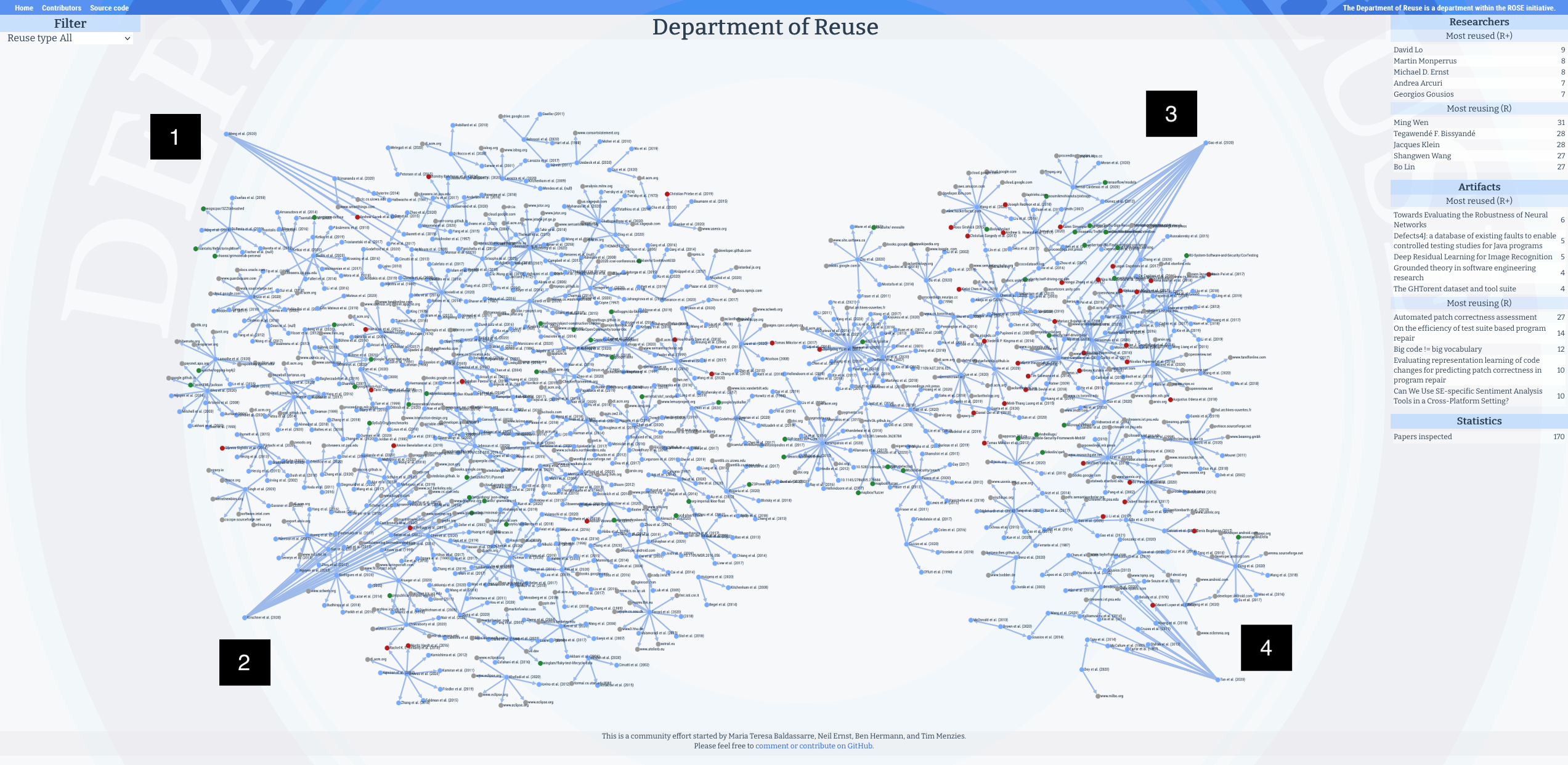}
\caption{From the web-site \url{https://reuse-dept.org}.
The  1,635 
arrows  in  this diagram connects reuser to reused. Blue dots denote the 714 sources found with with a digital object identifier (a DOI); e.g. any paper from a peer-reviewed
source. 
Red dots denote the 48 papers we found  without  DOIs  (e.g.  those  from arxiv.org). 
Gray and green denote
the 297  websites  and  57
Github repositories (respectively) reused in this sample. 
The black squares
pull out four examples where a paper has reused material from dozens of other
sources. Data collection for this graph is on-going and at the time 
of this writing, that data comes from 40\% of the papers published in the 2020 technical program of ICSE, ASE, FSE, ICMSE, MSR, and ESEM.}\label{reuse}
\end{figure}

We argue in this paper that, at least
in the  area  of
software engineering, this ``reuse problem'' is more apparent, than real.
We describe a successful experiment where teams of researchers   from
around the world read 170 recent (2020) conference papers from software engineering.
This work generated the ``reuse graph'' of Figure~\ref{reuse}.
In that  figure, each edge
connects papers to the prior work that they are (re-)using. 
As discussed below,
when compared to other  community
monitoring methods (e.g. artifact tracks or bibliometric  searches~\cite{Mathew18,baldassarre2019software}),
these reuse graphs
 are  less effort to build and  verify.
For example, it took around 12 minutes per paper for our team from
  Hong Kong, Canada, the United States, Italy,
Sweden, Finland, and Australia   to apply this reuse graph methodology to software engineering\footnote{That team included the authors of this paper
plus Jacky Keung from City University (Hong Kong); Greg
Gay from Chalmers University (Sweden);
Burak Turhan from Oulu University (Finland); and Aldeida Aleti from Monash University (Australia). We gratefully acknowledge their work, and that of their graduate
students.  In particular, we especially call out the work of Afonso Fontes
from Chalmers University (Sweden).}.

There are many other methods to map the structure of SE research
such as (a)~manual or automatic citations searchers 
or (b)~``artifact evaluation committees'' that aim to foster the generation
and sharing of research products (for more on artifact evaluation committees, see below).
Such studies  can lag significantly behind current work.
For example, our citation analysis of SE~\cite{Mathew18}   only went up
to 2016, the study itself was conducted in 2017, but not 
fully published till 2018. Given the enormous effort required for that work,
we have vowed never to do it again.
Reuse graphs, on the other hand,
are faster to keep up to date since the work of any one individual working on these
graphs is minimal.

Another reason to favor reuse graphs is that they
are community \textit{comprehensible}, community \textit{verifiable}, and  
community
\textit{correctable}.
All the data used for our reuse graphs is community-collected. All the data can be audited at \url{https://reuse-dept.org} and if errors are detected, issue reports can be raised in our GitHub repository (and the error  corrected). 
The same may {\em not} be true for  studies based on
citation servers run by professional bodies and for-profit organizations (e.g., see Table~\ref{errors}).

\begin{table}
\small
\begin{tabular}{|p{.95\linewidth}|}\hline
\rowcolor{blue!10}
{\bf EXAMPLE \#1:} At the time of this writing,  one of us has an entry in Google Scholar metrics software systems saying that their paper ``How to'' has 80 citations in the last five years at IEEE Transactions on Software Engineering. That link connects to some other paper, not written by any of us.  Yet there is no "help" button at Google Scholar where this error can be reported. 
 This is disappointing since
it is suspected that this mysterious ``How to'' paper references work that might be the most cited from IEEE TSE in the last five years.
That would be a significant achievement, if we could document it (but using
Google Scholar, we cannot).\\

{\bf EXAMPLE \#2:}
In our recent large scale text-mining studies of 30,000+ SE papers~\cite{Mathew18} ,
it was found that  papers can appear in our citation server, but not in others.
Also,  examples were found where some papers had twenty times the citations in one server
than in another. Here again, we were unable to contact anyone working on those citation servers to fix those errors. 
\\
\rowcolor{blue!10}
{\bf EXAMPLE \#3:}
Sometimes it is possible to contact the owners of these citations sites,
  but even then they may not fix errors. 
  For example, there  is no accepted convention for how to typeset a hypen, so different venues add in zero to one space before or after (and some even typeset the hyphen as two dashes). Hence, Zhou et al. found that  (a)~papers with hyphens in the title get reported as different papers in different venues; which means that (b)~those papers get fewer citations~\cite{Zhou21}.   Zhou et al. report that when they
  contacted
  the owners of these citation servers, rather than fix the errors,
  those owners started lobbying for the Zhou et al. paper not to be published. 
  \\\hline
  \end{tabular}
  
\caption{Examples of errors in  citation servers.}\label{errors}
  \end{table}


\respto{2.4} What is the value of a verified, continually updated,  snapshot of some current research
area? Once our reuse graph covers several years (and not just 2020 publications),
we foresee  several applications:
\begin{enumerate}
\item Graduate students could direct their attention to
research areas that are both very new (nodes from recent years) and very
productive (nodes with an unusually large number of edges attached);
\item The organizers of industrial and research conferences
could select their keynote speakers from that space
of new and productive artifacts.
\item
When applying for  promotion or hiring, research faculty or
industrial workers could document the impact of their work beyond papers, including tools, datasets, and innovative methods;
\item
Growth patterns   might
guide federal government funding priorities or departmental hiring plans.
\item
Venture capitalists could use these graphs to detect emergent technologies,
perhaps even funding some of those.
\item Conference organizers could check if their program committees
have enough members from currently hot topics.
\item Further, those same organizers could create new conference tracks and journals 
sections in order to service 
active research communities
that are under-represented in current publication
venues.
\item Journal editors could find reviewers with relevant experience. 
 \item Educators can use the graphs to guide their teaching plan.
 \end{enumerate}
 Further to the last point, we are planning an immediate
 application of Figure~\ref{reuse} graph for our   Fall'21 graduate SE classes. There,  we will tell students
that understanding the current state of the art will be their challenge for the rest of their career. But, using reuse graphs,   
it is possible for a community to find and maintain a
shared understanding of that state-of-the-art.
To demonstrate this,
in our Fall'21 classes, we are leaving the second half of the lecture plan blank. We will let students find and define
what cutting edge
techniques will be discussed there. 
To do so, their homework for the initial three weeks of class is to, at first,
learn this reuse graph approach 
by  performing  our standard ``reuse graph 101'' exercise\footnote{\url{https://github.com/bhermann/DoR/blob/main/workflow/training.md}}.
Then in week 2, read some papers to find their reuse (if any);
Finally,  in week 3,   they should  check someone else's reuse findings from other papers. 

 \section{Studying Reuse}
 
\begin{figure}
\includegraphics[width=6in]{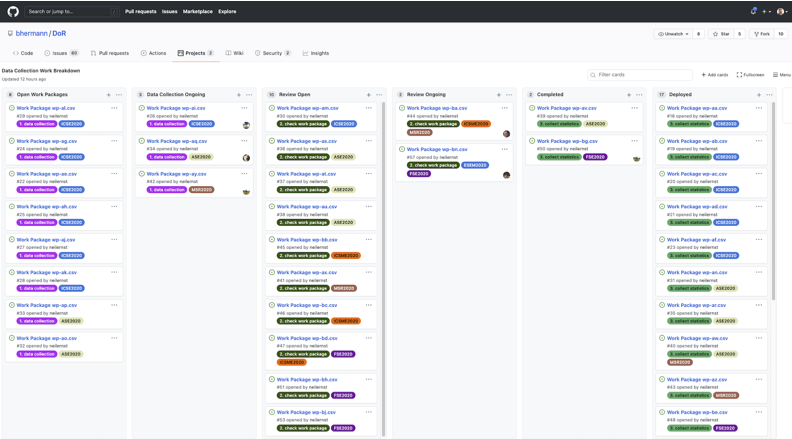}
\caption{Controlling data collection for building the reuse graphs.}\label{control}
\end{figure}

In our reuse study, we
  targeted   papers from the 2020 technical programs of six major international SE conferences: Software Engineering (ICSE), Automated Software Engineering (ASE), Joint European Software Engineering Conference / Foundations of Software Engineering (ESEC/FSE), Software Maintenance and Engineering (ICSME), Mining Software Repositories (MSR), and Empirical Software Engineering and Measurement (ESEM). 
These conferences
were selected using advice from~\cite{Mathew18}, but our vision is to expand; for example, by looking at all top-ranked SE conferences.
  GitHub issues were used to divide up the hundreds of papers
  from those conferences into ``work packets" of ten papers each.
  Reading teams were set up from software engineering research
teams from around the globe in Hong Kong, Istanbul (Turkey), Victoria (Canada),
  Gothenburg (Sweden), Oulu (Finland), Melbourne (Australia),  and Raleigh (USA). Team members would assign themselves work packets and
  then read the papers looking for the kinds of reuse enumerated below.
  Once completed, a second person (from any of our teams)
  would do the same and check for consistency. 
  Fleiss Kappa statistics are then computed
  to track the level of reader disagreement.
  All interaction was done via the GitHub issue system (see Figure~\ref{control}).
 
Teams were asked to record six kinds of reuse:
\begin{enumerate}
\item
Most papers have to benchmark their new ideas against some prior recent state-of-the-art paper. That is, they reuse old papers as {\em stepping stones} towards new results. 
\item
Another thing that is often reused are
statistical methods.  
Here we do not mean ``we use a two-tailed t-test'' or some other
decades-old widely-used statistical method.
Rather, we refer instead to  statistical methods for
 recent  papers that  propose
statistical guidance for the  kinds 
of analysis seen in SE . Perhaps because
this kind of analysis is very rare, these people
are exceedingly high cited; e.g. 

\bi
\item A 2008 paper 
{\em Benchmarking Classification Models for Software Defect Prediction}~\cite{lessmann08} currently
has 1,178 citations;
\item A 2011 paper {\em 
A practical guide for using statistical tests to assess randomized algorithms in software engineering}~\cite{acuri11} currently has 778 citations.
\ei 
\item
Metrics and Method descriptions
(which may be guidelines, with no tools);
\item
\respto{3.2a}\CHANGED{Data sets; }
\item
Sanity checks (justification for why a particular approach works or is reasonable to avoid bad data);
\item
and, indeed, the software packages of the kind currently
being reviewed by SE conference artifact evaluation committees (tools and replications).
\end{enumerate}

\begin{figure}
\begin{center}
    \begin{subfigure}{.32\textwidth}
    \centering
        \includegraphics[width=\linewidth]{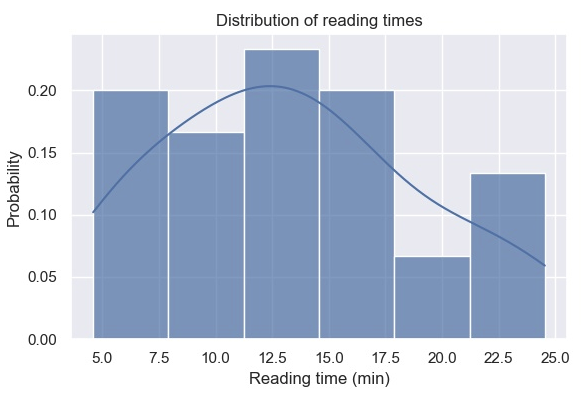}
        \caption{Reading times}
        \label{time}
    \end{subfigure}
    \begin{subfigure}{.32\textwidth}
    \centering
        \includegraphics[width=\linewidth]{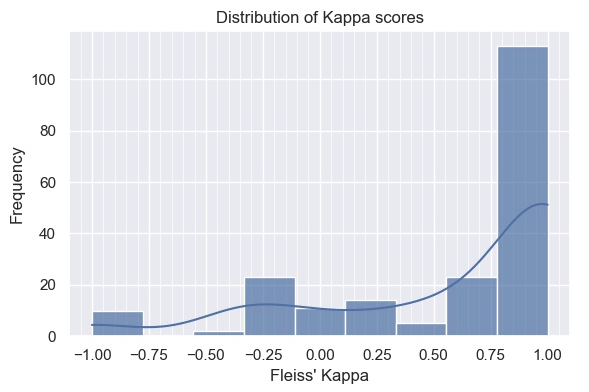}
        \caption{Fleiss Kappa scores}
        \label{agree}
    \end{subfigure}
    \begin{subfigure}{.32\textwidth}
    \centering
        \includegraphics[width=\linewidth]{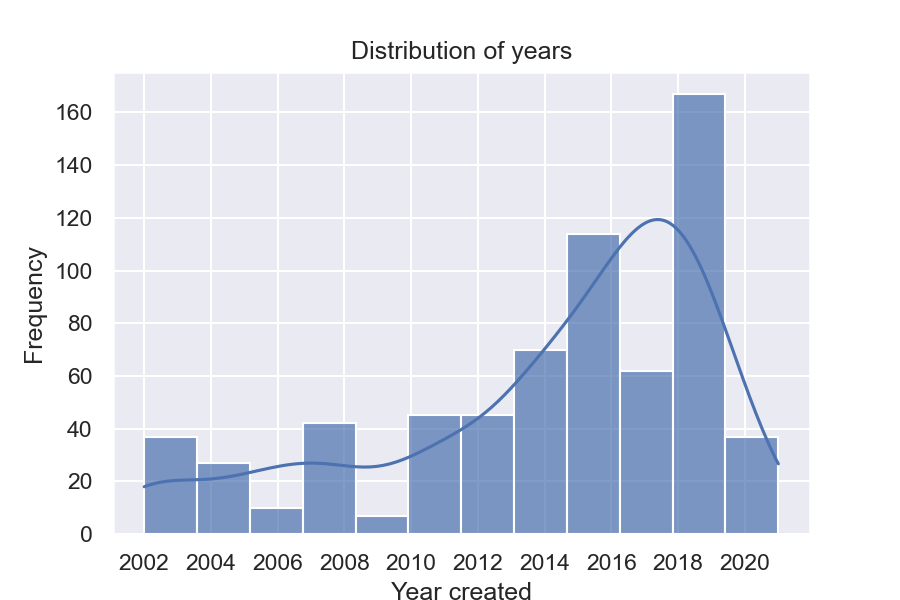}
        \caption{Years that the works were created.\respto{3.1}.}
        \label{years}
    \end{subfigure}
\end{center}
    \caption{Reading time results (left). Agreement scores (center). Yearly prevalence of reused papers (right).}\label{results}
\end{figure}

We can report that it is not difficult to read papers in order
to detect these kinds of reuse:
\bi
\item It is fast to find the above six kinds of reuse. Our graduate
students report that reading their first paper might take up to an
hour. But after two or three papers, the median reading time drops to around 12 minutes
(see Figure~\ref{time}).
\item
When we compare the reuse reported by different readers,
we get Figure~\ref{agree}.
In our current results, the median   Fleiss Kappa score
(for reviewer agreement) is one (i.e. very good).
\ei
Of course, there any many other items being reused than the six listed above\footnote{See https://pasteboard.co/Ke4tKgO.png}.  It is an open question,
worthy of future work, to check if those other items can be collected in this way.

\section{Related Work}

Apart from software engineering, 
many other disciplines are actively engaged in artifact creation, sharing, and re-use.
Artifacts are useful for building a culture of replication and reproducibility, already acknowledged as important in SE \cite{2020arXiv201102861S}. Fields such as psychology have had many early results thrown into doubt because of a failure to replicate the original findings~\cite{Schimmack2020}. Sharing research protocols and data allows for other research teams to conduct \textit{severe} tests of the original studies \cite{Mayo2018}, strengthening (or rejecting) these initial findings. In medicine, drug companies are mandated to share the research protocols and outcomes of their drug trials, something that has become of vital recent importance (albeit not without challenges \cite{DeVito2020}). In physics and astronomy, artifact sharing is so commonplace that large community infrastructures exist solely to ensure data sharing, not least because governments which fund these costly experiments insist on it. 

Furthermore, software plays such a vital role in this enterprise that many fields have begun training and developing science-specific research software engineers, for example, at the UK's Software Sustainability Institute\footnote{https://www.software.ac.uk}, or the USA's NSF-funded Molecular Science Software Institute\footnote{https://molssi.org}. Indeed, a founding impetus for the World Wide Web was the need for CERN (Center for Nuclear Research) to facilitate knowledge sharing~\cite{tbl90}\footnote{And Tim Berners-Lee himself has said ``Had the technology been proprietary, and in my total control, it would probably not have taken off.'' Another argument for artifacts!}---a mission now continued by Zenodo, also operated by CERN. 

In more theoretical areas of CS, pioneering use of preprint servers has enabled `reuse' of proofs, essential to progress. 
In machine learning, replication is focused on stepping-stones, enabled by highly successful benchmarks such as ImageNet \cite{ILSVRC15}. However, recent advances with extremely costly training regimens have called replicability into question\footnote{\url{https://www.technologyreview.com/2020/11/12/1011944/artificial-intelligence-replication-crisis-science-big-tech-google-deepmind-facebook-openai/}}.

In the specific case of software engineering research,
prior to this paper, there was little recorded and verified evidence
of reuse.
Many researchers have conducted {\em citation studies} that
find links to highly cited papers (e.g.~\cite{Mathew18}).
 As stated in our introduction,
  such studies can lag behind the latest results.
Also, recalling Table~\ref{errors}, we have 
cause to doubt the conclusions from such citation studies.

\subsection{What about Artifact Evaluation Committees?}\label{aec}
 \respto{1.1}  Another practice that is becoming increasingly common
  is for conferences to run  {\em
  artifact evaluation committees}.
The authors of accepted conference papers  submit software packages  that, in theory,
let others  re-execute that work.
These evaluation committees award ``badges'' as shown in Table~\ref{badges}.

\begin{table*}[!bt]
    \caption{Badges currently awarded at ACM conferences~\cite{ACMguidelines}. 
This table is for the ACM  and analogous tables are used at other conferences.} \label{badges}
{\scriptsize
    \begin{center}
    \begin{tabular}{p{2.5cm}p{2.5cm}p{2.5cm}p{2.5cm}p{2.5cm}}
      {\centering \textbf{Available}\par}  &  {\centering \textbf{Functional}\par} & {\centering \textbf{Reusable}\par}& {\centering \textbf{Reproduced}\par} & {\centering \textbf{Replicated}\par} \\
       {\centering \includegraphics[width=1.5cm]{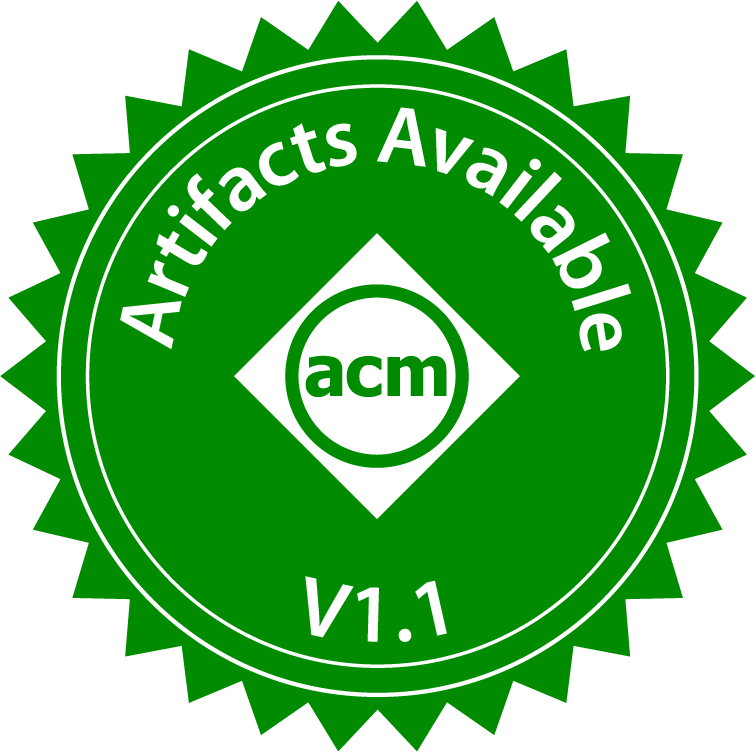}\par} &  {\centering \includegraphics[width=1.5cm]{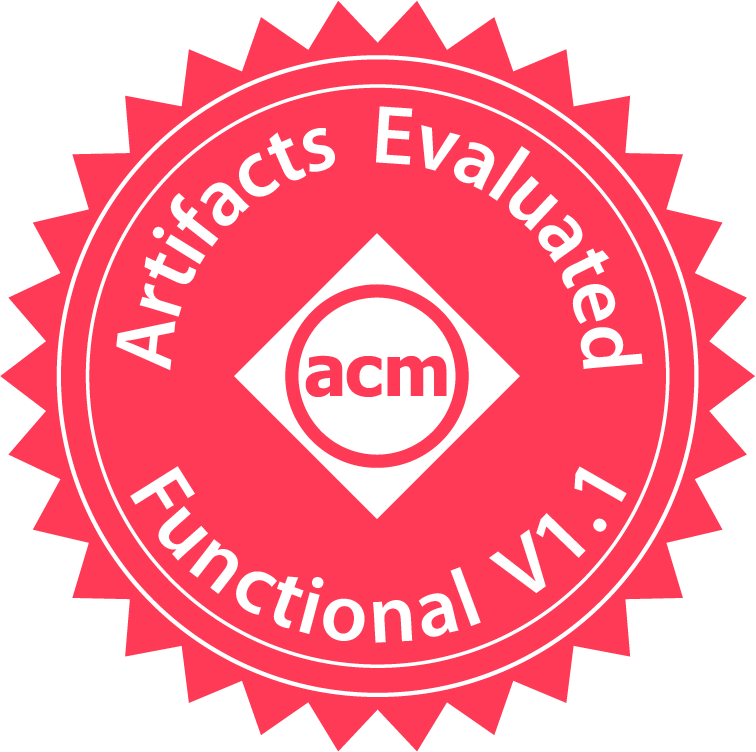}\par}& {\centering \includegraphics[width=1.5cm]{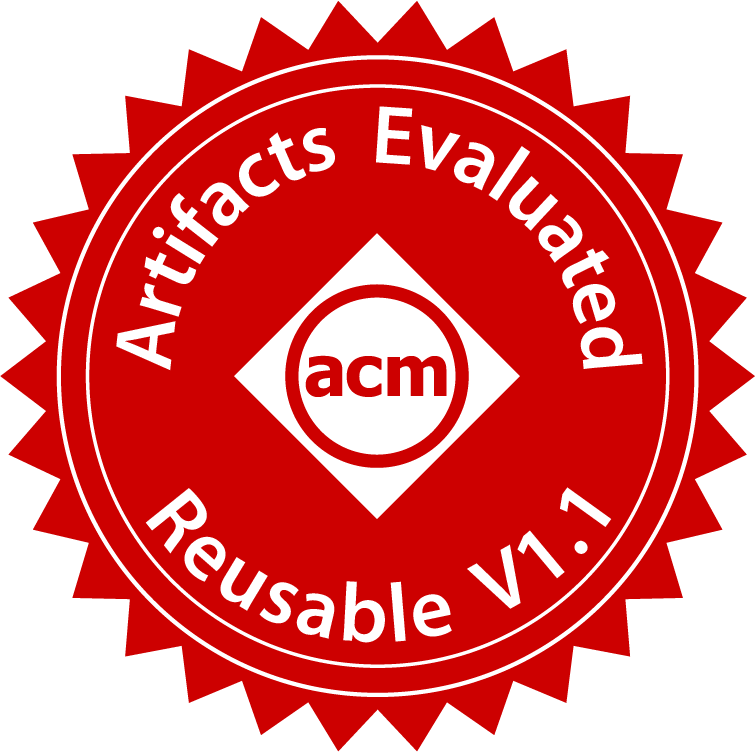}\par} & {\centering \includegraphics[width=1.5cm]{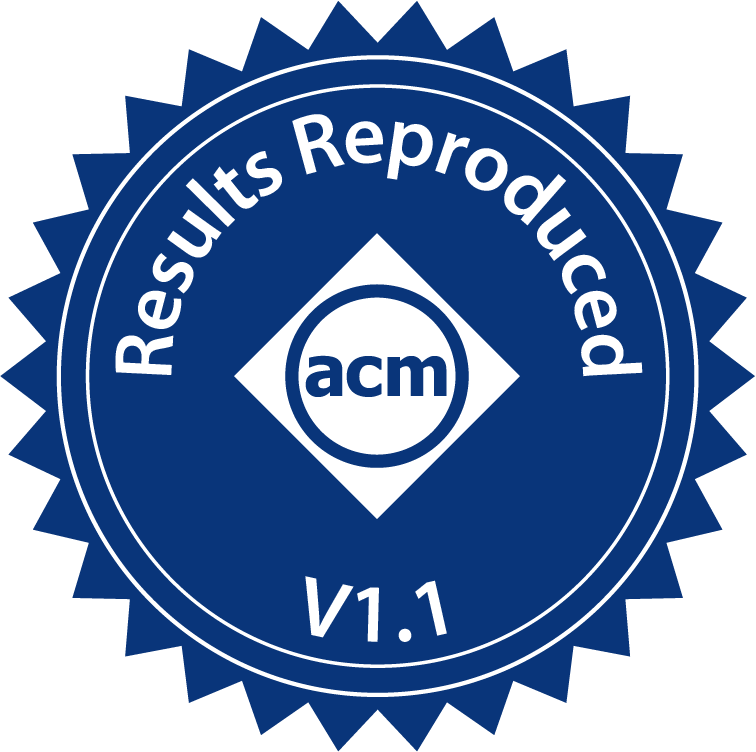}\par} & {\centering \includegraphics[width=1.5cm]{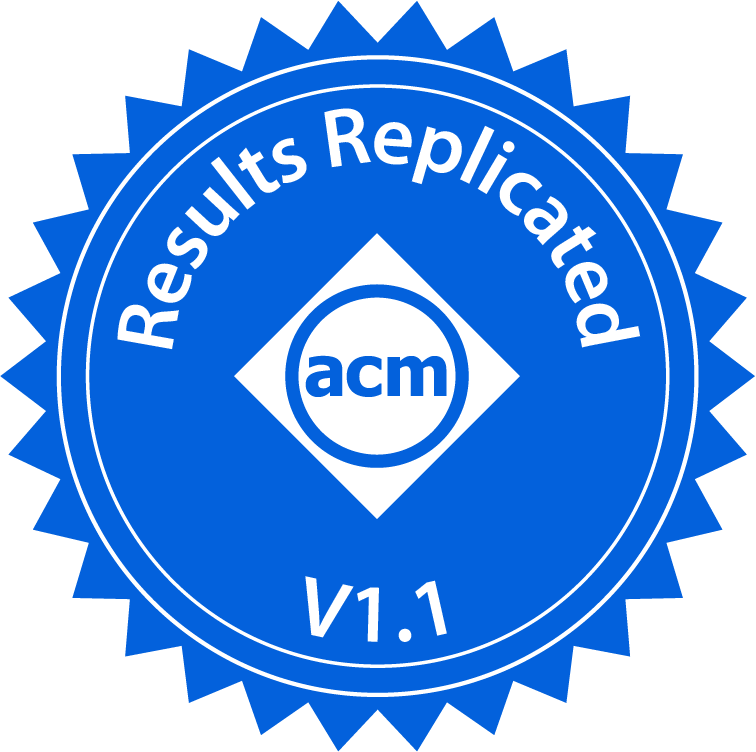}\par}  \\
       Archived in a public repository with a long-term retention policy. A DOI needs to be provided. & Artifacts need to be documented, consistent, complete, exercisable, and include appropriate evidence of verification and validation. & Only available to artifacts already qualifying for the functional badge. Needs to significantly exceed minimal functionality. & Results of this paper have been reproduced by a different team using the original artifact. & Results of this paper have been replicated by a different team without the original artifact. 
    \end{tabular}
    \end{center}}

\end{table*} 

 \begin{figure}[bt]
    \centering
    \includegraphics[width=.75\textwidth]{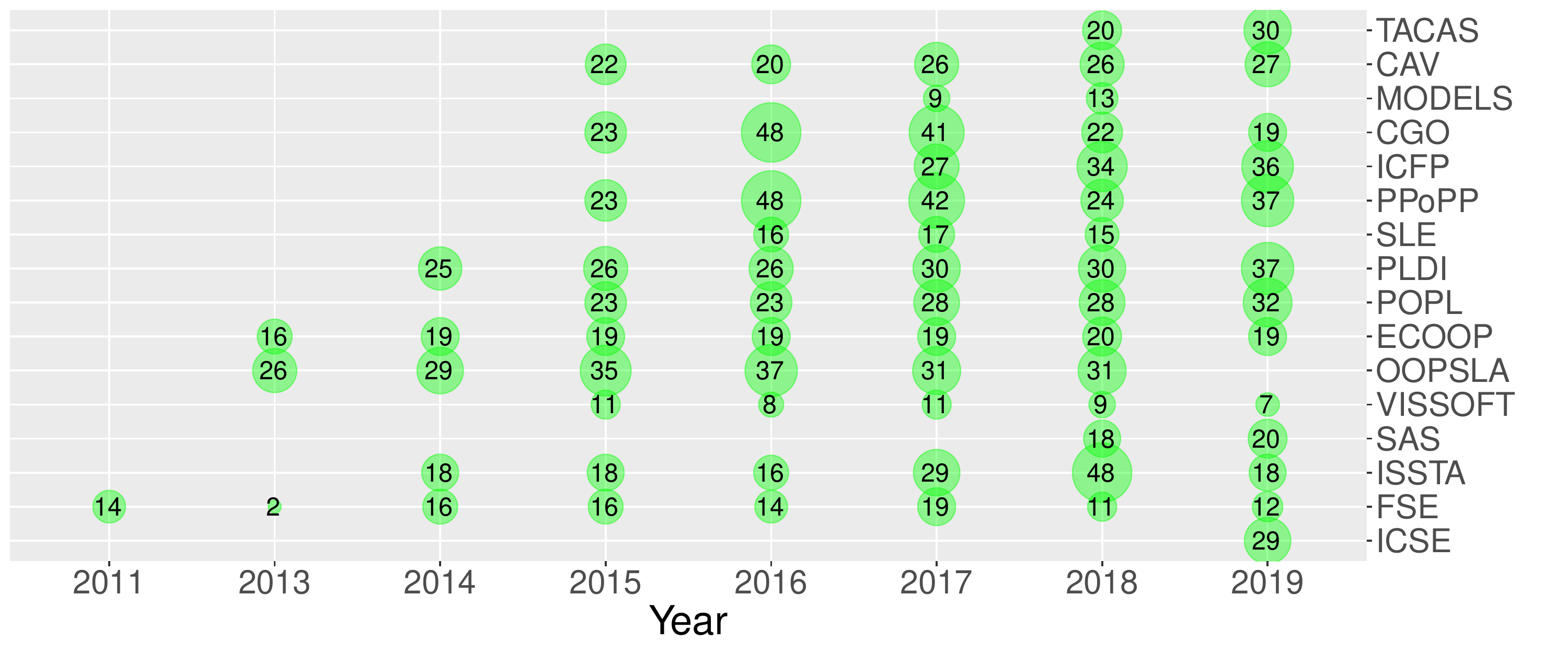}
    \caption{Artifact evaluation committee sizes 2011-2019.
    From  Hermann et al.~\cite{10.1145/3368089.3409767}}
    \label{fig:committee_sizes}
\end{figure}

 

Artifact evaluation is something of a ``growth industry''
in SE (and programming languages, or PL, community).
Figure~\ref{fig:committee_sizes} shows the increasing number of people
 evaluating artifacts 2011 to 2019. As to more recent data:
 (a)~at PLDI'20, 61 of 77 papers offered artifacts~\cite{10.1145/3385412}, and at ECOOP'21, all 20 research papers offered software artifacts~\cite{mller_et_al:LIPIcs.ECOOP.2021.0};
 (b)~at ASE'21, they have a 60 person artifact evaluation track.

The question has to be asked:
are all the people of  Figure~\ref{fig:committee_sizes} making the best
use of their time?
Perhaps not.
 We note that most artifacts are assigned the badges
 requested by the authors. Given that, it might be safe
 to ask some of the personnel 
 from  Figure~\ref{fig:committee_sizes} to (e.g.)
 spend less time on evaluating conference artifacts
 and spend more time working on
 Figure~\ref{reuse}.

  Also, we suspect that the badges of Table~\ref{badges}
need refinement since much time can be squandered
on   minor issues with little practical effect.
 For example, it can be hard to distinguish ``functional'' from ``reusable'' (in fact, some artifact evaluation committees just ignore the ``functional'' badge; e.g. see the artifact evaluation process at ICSE'21, ICMSE'21, and ASE'21).
 
Further, as to  checking for ``functional'' and ``reusable'', that
 requires  downloading
 then installing then  running the software. This can be a very long process (taking hours to days), especially for quirky research prototypes where (e.g.) the   scripts have one letter typos and/or
 the install instructions are missing small, but crucial pieces of information.

But most importantly, it is not clear that the artifact
 evaluation process is creating {\em reused} artifacts.
 If we query ACM Portal for ``software engineering" and ``artifacts" in the range 2015 to 2020, we find that most of the recorded artifacts are 
 \respto{1.4}  not reused in replications or reproductions\footnote{ As of December 10 2020, that search returns 2,535 SE papers  with artifact badge.
 Of these,  43\%, 30\%, 20\%, 5\%, 2\% are  available, functional, reusable, reproduced and replicated artifacts (respectively).}.
Specifically, only $\frac{1}{20}$th are reproduced and only
 $\frac{1}{50}$th are replicated.  
  \respto{1.2}
One possibility for these results, which we can quickly discount, is
that  SE artifacts are a new idea that will take a while to catch on.
If this were the case, we would expect that older SE artifacts are reproduced  more, since they have been around  for longer.
But this is not the case. Look at Figure~\ref{results}c,
we see that most of the reused artifacts were created
very recently.

Perhaps we need to change the definition
of the badges and say an artifact is ``reusable'' {\em if it is reused} (and not before).
Also, it might be useful to reflect more on what is actually being reused (as we have done, above).

\section{Next Steps for Reuse Graphs}

While
Figure~\ref{reuse} is a promising start, to scale up from here,  we need to organize a larger
reading population.
Our goal  is to analyze 200, 2000, 5000 papers
in 2021, 2022, 2023 (respectively)
by which time we would have covered
most of the major SE conferences in the last five years.  
After that, our maintenance goal would be to read 500 (ish)
papers per year to keep up to date with the conferences. 
 \respto{3.6}
 Based on Figure~\ref{results}a, then assuming each paper is read by two people,  then maintenance goal would be achievable by a team of twenty people working two hours per month on this task.
To organize this work, we have created
the ``ROSE initiative'', see Table~\ref{rose}.
\begin{table}
\caption{The ROSE initiative: \underline{R}ecognizing and Rewarding 
\underline{O}pen Science in \underline{S}oftware \underline{E}ngineering:
 an international  multi-conference workshop that will continually
report updates to the SE reuse  graphs.}\label{rose}
\begin{tabular}{|p{.98\linewidth}|}\hline
\rowcolor{blue!10}
Researchers that reuse the most from other papers will be applauded and awarded a ``R-index'' (reuse index). \\
Researchers that build the artifacts that are most reused will be applauded
(even louder) and be awarded an ``R+-index'' indicating that they are the people
producing the artifacts that are most used by the rest of the community.
\\\rowcolor{blue!10}
In between each conference, the ROSE initiative will co-ordinate an international
team of volunteers incrementally updating the SE reuse graph. 
\\
This reuse
graph will be displayed at a publicly available web site (reuse-dept.org).
Individual researchers can browse that site, check their entries, and propose corrections and extensions.
\\\rowcolor{blue!10}
All reports of reuse will be double checked and  disputed
claims will be then be tripled-checked.
\\
All the tools used to create that web site will be freely available for download. Hence, if the SE community does not like how we are running these reuse graphs, they can take all our code and data and do something else. \\\rowcolor{blue!10}
Also, researchers from other disciplines can take
our tools and apply them  to their own community.\\\hline
\end{tabular}
\end{table}
 If that work interests you, then there are many ways you can get involved:
\bi
\item
If you are a researcher and wish to check that we have accurately recorded your contribution,
please visit \url{https://reuse-dept.org}.  
\item
If you want to apply reuse graphs
to your community, please use our tools at \url{https://github.com/bhermann/DoR/}.
 \item
 If you are interested in joining this initiative and contributing to an up-to-minute snapshot of SE research, then please 
(a)~take our how-to-read-for-reuse tutorial\footnote{https://github.com/bhermann/DoR/blob/main/workflow/training.md};
(b)~then visit the dashboard at Figure~\ref{control},
find an issue with no one's face on it, and assign yourself a task.
\item
Better yet, if you are an educator teaching a graduate SE
class, then get your students to do the three week reading assignments shown in the introduction. As a result, students will join an international team exploring reuse in SE that will keep them informed and updated about the state-of-the-art in SE for many years to come. Also, as a side-effect, they will also see first hand the benefit of open source tools that can be shared by teams working around the globe.
\ei

We see this effort as one part of the broader open science effort, in addition to helping the community identify the state of the art (e.g., patterns of growth in the reuse graph). Among the goals of open science are to increase confidence in published results, and to acknowledge that science produces more types of artifacts than just publications: researchers also produce method innovations, new datasets, and improved tools. If we take an agile view of SE science, then as researchers we should focus on generating these artifacts and rapidly securing critique, curation, and clarification from our peers and the public.
  
\bibliographystyle{ACM-Reference-Format}      
\bibliography{refs,refs-biblio.bib}


\begin{thebibliography}{17}


\ifx \showCODEN    \undefined \def \showCODEN     #1{\unskip}     \fi
\ifx \showDOI      \undefined \def \showDOI       #1{#1}\fi
\ifx \showISBNx    \undefined \def \showISBNx     #1{\unskip}     \fi
\ifx \showISBNxiii \undefined \def \showISBNxiii  #1{\unskip}     \fi
\ifx \showISSN     \undefined \def \showISSN      #1{\unskip}     \fi
\ifx \showLCCN     \undefined \def \showLCCN      #1{\unskip}     \fi
\ifx \shownote     \undefined \def \shownote      #1{#1}          \fi
\ifx \showarticletitle \undefined \def \showarticletitle #1{#1}   \fi
\ifx \showURL      \undefined \def \showURL       {\relax}        \fi
\providecommand\bibfield[2]{#2}
\providecommand\bibinfo[2]{#2}
\providecommand\natexlab[1]{#1}
\providecommand\showeprint[2][]{arXiv:#2}

\bibitem[\protect\citeauthoryear{Arcuri and Briand}{Arcuri and Briand}{2011}]%
        {acuri11}
\bibfield{author}{\bibinfo{person}{Andrea Arcuri} {and} \bibinfo{person}{Lionel
  Briand}.} \bibinfo{year}{2011}\natexlab{}.
\newblock \showarticletitle{A Practical Guide for Using Statistical Tests to
  Assess Randomized Algorithms in Software Engineering}. In
  \bibinfo{booktitle}{\emph{Proceedings of the 33rd International Conference on
  Software Engineering}} (Waikiki, Honolulu, HI, USA)
  \emph{(\bibinfo{series}{ICSE '11})}. \bibinfo{publisher}{Association for
  Computing Machinery}, \bibinfo{address}{New York, NY, USA},
  \bibinfo{pages}{1–10}.
\newblock
\showISBNx{9781450304450}
\urldef\tempurl%
\url{https://doi.org/10.1145/1985793.1985795}
\showDOI{\tempurl}


\bibitem[\protect\citeauthoryear{{Association for Computing
  Machinery}}{{Association for Computing Machinery}}{2020}]%
        {ACMguidelines}
\bibfield{author}{\bibinfo{person}{{Association for Computing Machinery}}.}
  \bibinfo{year}{2020}\natexlab{}.
\newblock \bibinfo{title}{{Artifact Review and Badging}}.
\newblock
  \bibinfo{howpublished}{\url{https://www.acm.org/publications/policies/artifact-review-and-badging-current}}.
\newblock
\newblock
\shownote{Accessed: 2020-12-08.}


\bibitem[\protect\citeauthoryear{Baldassarre, Caivano, Romano, and
  Scanniello}{Baldassarre et~al\mbox{.}}{2019}]%
        {baldassarre2019software}
\bibfield{author}{\bibinfo{person}{Maria~Teresa Baldassarre},
  \bibinfo{person}{Danilo Caivano}, \bibinfo{person}{Simone Romano}, {and}
  \bibinfo{person}{Giuseppe Scanniello}.} \bibinfo{year}{2019}\natexlab{}.
\newblock \showarticletitle{Software models for source code maintainability: A
  systematic literature review}. In \bibinfo{booktitle}{\emph{2019 45th
  Euromicro Conference on Software Engineering and Advanced Applications
  (SEAA)}}. IEEE, \bibinfo{pages}{252--259}.
\newblock


\bibitem[\protect\citeauthoryear{Berners-Lee}{Berners-Lee}{1990}]%
        {tbl90}
\bibfield{author}{\bibinfo{person}{Tim Berners-Lee}.}
  \bibinfo{year}{1990}\natexlab{}.
\newblock \bibinfo{booktitle}{\emph{Information Management: A Proposal}}.
\newblock
\urldef\tempurl%
\url{https://www.w3.org/History/1989/proposal.html}
\showURL{%
\tempurl}


\bibitem[\protect\citeauthoryear{da~Silva, Suassuna, Fran{\c{c}}a, Grubb,
  Gouveia, Monteiro, and dos Santos}{da~Silva et~al\mbox{.}}{2012}]%
        {daSilva2012}
\bibfield{author}{\bibinfo{person}{Fabio Q.~B. da Silva},
  \bibinfo{person}{Marcos Suassuna}, \bibinfo{person}{A.~C{\'{e}}sar~C.
  Fran{\c{c}}a}, \bibinfo{person}{Alicia~M. Grubb}, \bibinfo{person}{Tatiana~B.
  Gouveia}, \bibinfo{person}{Cleviton V.~F. Monteiro}, {and}
  \bibinfo{person}{Igor~Ebrahim dos Santos}.} \bibinfo{year}{2012}\natexlab{}.
\newblock \showarticletitle{Replication of empirical studies in software
  engineering research: a systematic mapping study}.
\newblock \bibinfo{journal}{\emph{Empirical Software Engineering}}
  (\bibinfo{date}{Sept.} \bibinfo{year}{2012}).
\newblock
\urldef\tempurl%
\url{https://doi.org/10.1007/s10664-012-9227-7}
\showDOI{\tempurl}


\bibitem[\protect\citeauthoryear{DeVito, Bacon, and Goldacre}{DeVito
  et~al\mbox{.}}{2020}]%
        {DeVito2020}
\bibfield{author}{\bibinfo{person}{Nicholas~J DeVito}, \bibinfo{person}{Seb
  Bacon}, {and} \bibinfo{person}{Ben Goldacre}.}
  \bibinfo{year}{2020}\natexlab{}.
\newblock \showarticletitle{Compliance with legal requirement to report
  clinical trial results on {ClinicalTrials}.gov: a cohort study}.
\newblock \bibinfo{journal}{\emph{The Lancet}} \bibinfo{volume}{395},
  \bibinfo{number}{10221} (\bibinfo{date}{feb} \bibinfo{year}{2020}),
  \bibinfo{pages}{361--369}.
\newblock
\urldef\tempurl%
\url{https://doi.org/10.1016/s0140-6736(19)33220-9}
\showDOI{\tempurl}


\bibitem[\protect\citeauthoryear{Donaldson and Torlak}{Donaldson and
  Torlak}{2020}]%
        {10.1145/3385412}
\bibfield{editor}{\bibinfo{person}{Alastair~F. Donaldson} {and}
  \bibinfo{person}{Emina Torlak}} (Eds.). \bibinfo{year}{2020}\natexlab{}.
\newblock \bibinfo{booktitle}{\emph{PLDI 2020: Proceedings of the 41st ACM
  SIGPLAN Conference on Programming Language Design and Implementation}}
  (London, UK). \bibinfo{publisher}{Association for Computing Machinery},
  \bibinfo{address}{New York, NY, USA}.
\newblock
\showISBNx{9781450376136}
\urldef\tempurl%
\url{https://doi.org/10.1145/3385412}
\showDOI{\tempurl}


\bibitem[\protect\citeauthoryear{Hermann, Winter, and Siegmund}{Hermann
  et~al\mbox{.}}{2020}]%
        {10.1145/3368089.3409767}
\bibfield{author}{\bibinfo{person}{Ben Hermann}, \bibinfo{person}{Stefan
  Winter}, {and} \bibinfo{person}{Janet Siegmund}.}
  \bibinfo{year}{2020}\natexlab{}.
\newblock \showarticletitle{Community Expectations for Research Artifacts and
  Evaluation Processes}. In \bibinfo{booktitle}{\emph{Proceedings of the 28th
  ACM Joint Meeting on European Software Engineering Conference and Symposium
  on the Foundations of Software Engineering}} (Virtual Event, USA)
  \emph{(\bibinfo{series}{ESEC/FSE 2020})}. \bibinfo{publisher}{Association for
  Computing Machinery}, \bibinfo{address}{New York, NY, USA},
  \bibinfo{pages}{469–480}.
\newblock
\showISBNx{9781450370431}
\urldef\tempurl%
\url{https://doi.org/10.1145/3368089.3409767}
\showDOI{\tempurl}


\bibitem[\protect\citeauthoryear{Lessmann, Baesens, Mues, and Pietsch}{Lessmann
  et~al\mbox{.}}{2008}]%
        {lessmann08}
\bibfield{author}{\bibinfo{person}{Stefan Lessmann}, \bibinfo{person}{Bart
  Baesens}, \bibinfo{person}{Christophe Mues}, {and} \bibinfo{person}{Swantje
  Pietsch}.} \bibinfo{year}{2008}\natexlab{}.
\newblock \showarticletitle{Benchmarking Classification Models for Software
  Defect Prediction: A Proposed Framework and Novel Findings}.
\newblock \bibinfo{journal}{\emph{IEEE Transactions on Software Engineering}}
  \bibinfo{volume}{34}, \bibinfo{number}{4} (\bibinfo{year}{2008}),
  \bibinfo{pages}{485--496}.
\newblock
\urldef\tempurl%
\url{https://doi.org/10.1109/TSE.2008.35}
\showDOI{\tempurl}


\bibitem[\protect\citeauthoryear{Mathew, Agrawal, and Menzies}{Mathew
  et~al\mbox{.}}{2018}]%
        {Mathew18}
\bibfield{author}{\bibinfo{person}{George Mathew}, \bibinfo{person}{Amritanshu
  Agrawal}, {and} \bibinfo{person}{Tim Menzies}.}
  \bibinfo{year}{2018}\natexlab{}.
\newblock \showarticletitle{Finding Trends in Software Research}.
\newblock \bibinfo{journal}{\emph{IEEE Transactions on Software Engineering}}
  (\bibinfo{year}{2018}), \bibinfo{pages}{1--1}.
\newblock
\urldef\tempurl%
\url{https://doi.org/10.1109/TSE.2018.2870388}
\showDOI{\tempurl}


\bibitem[\protect\citeauthoryear{Mayo}{Mayo}{2018}]%
        {Mayo2018}
\bibfield{author}{\bibinfo{person}{D.~G. Mayo}.}
  \bibinfo{year}{2018}\natexlab{}.
\newblock \bibinfo{booktitle}{\emph{Statistical inference as severe testing:
  How to get beyond the statistics wars}}.
\newblock \bibinfo{publisher}{Cambridge University Press}.
\newblock


\bibitem[\protect\citeauthoryear{M{\o}ller and Sridharan}{M{\o}ller and
  Sridharan}{2021}]%
        {mller_et_al:LIPIcs.ECOOP.2021.0}
\bibfield{author}{\bibinfo{person}{Anders M{\o}ller} {and}
  \bibinfo{person}{Manu Sridharan}.} \bibinfo{year}{2021}\natexlab{}.
\newblock \showarticletitle{{Front Matter, Table of Contents, Preface,
  Conference Organization}}. In \bibinfo{booktitle}{\emph{35th European
  Conference on Object-Oriented Programming (ECOOP 2021)}}
  \emph{(\bibinfo{series}{Leibniz International Proceedings in Informatics
  (LIPIcs)}, Vol.~\bibinfo{volume}{194})},
  \bibfield{editor}{\bibinfo{person}{Anders M{\o}ller} {and}
  \bibinfo{person}{Manu Sridharan}} (Eds.). \bibinfo{publisher}{Schloss
  Dagstuhl -- Leibniz-Zentrum f{\"u}r Informatik}, \bibinfo{address}{Dagstuhl,
  Germany}, \bibinfo{pages}{0:i--0:xxiv}.
\newblock
\showISBNx{978-3-95977-190-0}
\showISSN{1868-8969}
\urldef\tempurl%
\url{https://doi.org/10.4230/LIPIcs.ECOOP.2021.0}
\showDOI{\tempurl}


\bibitem[\protect\citeauthoryear{Popper}{Popper}{2014}]%
        {popper2014conjectures}
\bibfield{author}{\bibinfo{person}{Karl Popper}.}
  \bibinfo{year}{2014}\natexlab{}.
\newblock \bibinfo{booktitle}{\emph{Conjectures and refutations: The growth of
  scientific knowledge}}.
\newblock \bibinfo{publisher}{Routledge}.
\newblock


\bibitem[\protect\citeauthoryear{Russakovsky, Deng, Su, Krause, Satheesh, Ma,
  Huang, Karpathy, Khosla, Bernstein, Berg, and Fei-Fei}{Russakovsky
  et~al\mbox{.}}{2015}]%
        {ILSVRC15}
\bibfield{author}{\bibinfo{person}{Olga Russakovsky}, \bibinfo{person}{Jia
  Deng}, \bibinfo{person}{Hao Su}, \bibinfo{person}{Jonathan Krause},
  \bibinfo{person}{Sanjeev Satheesh}, \bibinfo{person}{Sean Ma},
  \bibinfo{person}{Zhiheng Huang}, \bibinfo{person}{Andrej Karpathy},
  \bibinfo{person}{Aditya Khosla}, \bibinfo{person}{Michael Bernstein},
  \bibinfo{person}{Alexander~C. Berg}, {and} \bibinfo{person}{Li Fei-Fei}.}
  \bibinfo{year}{2015}\natexlab{}.
\newblock \showarticletitle{{ImageNet Large Scale Visual Recognition
  Challenge}}.
\newblock \bibinfo{journal}{\emph{International Journal of Computer Vision
  (IJCV)}} \bibinfo{volume}{115}, \bibinfo{number}{3} (\bibinfo{year}{2015}),
  \bibinfo{pages}{211--252}.
\newblock
\urldef\tempurl%
\url{https://doi.org/10.1007/s11263-015-0816-y}
\showDOI{\tempurl}


\bibitem[\protect\citeauthoryear{Santos, Vegas, Oivo, and Juristo}{Santos
  et~al\mbox{.}}{2021}]%
        {2020arXiv201102861S}
\bibfield{author}{\bibinfo{person}{Adrian Santos}, \bibinfo{person}{Sira
  Vegas}, \bibinfo{person}{Markku Oivo}, {and} \bibinfo{person}{Natalia
  Juristo}.} \bibinfo{year}{2021}\natexlab{}.
\newblock \showarticletitle{Comparing the results of replications in software
  engineering}.
\newblock \bibinfo{journal}{\emph{Empirical Software Engineering}}
  \bibinfo{volume}{26}, \bibinfo{number}{2} (\bibinfo{date}{Feb.}
  \bibinfo{year}{2021}).
\newblock
\urldef\tempurl%
\url{https://doi.org/10.1007/s10664-020-09907-7}
\showDOI{\tempurl}


\bibitem[\protect\citeauthoryear{Schimmack}{Schimmack}{2020}]%
        {Schimmack2020}
\bibfield{author}{\bibinfo{person}{Ulrich Schimmack}.}
  \bibinfo{year}{2020}\natexlab{}.
\newblock \showarticletitle{A meta-psychological perspective on the decade of
  replication failures in social psychology.}
\newblock \bibinfo{journal}{\emph{Canadian Psychology/Psychologie canadienne}}
  \bibinfo{volume}{61}, \bibinfo{number}{4} (\bibinfo{date}{nov}
  \bibinfo{year}{2020}), \bibinfo{pages}{364--376}.
\newblock
\urldef\tempurl%
\url{https://doi.org/10.1037/cap0000246}
\showDOI{\tempurl}


\bibitem[\protect\citeauthoryear{Zhou, Tse, and Witheridge}{Zhou
  et~al\mbox{.}}{2021}]%
        {Zhou21}
\bibfield{author}{\bibinfo{person}{Zhi~Quan Zhou}, \bibinfo{person}{T.~H. Tse},
  {and} \bibinfo{person}{Matt Witheridge}.} \bibinfo{year}{2021}\natexlab{}.
\newblock \showarticletitle{Metamorphic Robustness Testing: Exposing Hidden
  Defects in Citation Statistics and Journal Impact Factors}.
\newblock \bibinfo{journal}{\emph{IEEE Transactions on Software Engineering}}
  \bibinfo{volume}{47}, \bibinfo{number}{6} (\bibinfo{year}{2021}),
  \bibinfo{pages}{1164--1183}.
\newblock
\urldef\tempurl%
\url{https://doi.org/10.1109/TSE.2019.2915065}
\showDOI{\tempurl}


\end{thebibliography}
\end{document}